\documentclass[groupedaddress,amsmath,amssymb,aps,prd,twocolumn,nofootinbib,nobibnotes,notitlepage,floatfix,superscriptaddress,reprint]{revtex4-2}

\usepackage{graphicx}
\usepackage{dcolumn}
\usepackage{bm}
\usepackage{multirow,booktabs}
\usepackage[table]{xcolor}
\usepackage{enumitem}
\usepackage{mathtools}
\usepackage{caption}
\usepackage{calc}
\usepackage{cancel}
\usepackage{framed}
\usepackage{subcaption}
\usepackage{ragged2e}
\usepackage{float}
\usepackage{soul}
\usepackage[normalem]{ulem}  
\usepackage{color} 
\renewcommand{\sout}{\bgroup \color{red} \ULdepth=-.5ex \ULset}

\newcommand{\Tr}{\text{Tr}}

\newcommand{\rot}{{\rm{rot}}}

\newcommand{\VEV}[1]{\langle {{#1}}\rangle}

\begin{document}

\title{\textbf{Proton Spin Decomposition via QCD Sum Rules}}
\author{HyungJoo Kim}
 \email{hugokm0322@gmail.com}
\affiliation{International Institute for Sustainability with Knotted Chiral Meta Matter (WPI-SKCM$^2$), Hiroshima University, Higashi-Hiroshima, Hiroshima 739-8526, Japan}
\author{Philipp Gubler}%
\email{philipp.gubler1@gmail.com}
\affiliation{Advanced Science Research Center, Japan Atomic Energy Agency, Tokai, Ibaraki 319-1195, Japan}%
\author{Chihiro Sasaki}
\email{chihiro.sasaki@uwr.edu.pl}
\affiliation{Institute of Theoretical Physics, University of Wroc\l{}aw, plac Maksa Borna 9, PL-50204 Wroc\l{}aw, Poland}
\affiliation{International Institute for Sustainability with Knotted Chiral Meta Matter (WPI-SKCM$^2$), Hiroshima University, Higashi-Hiroshima, Hiroshima 739-8526, Japan}

\date{\today}

\begin{abstract}
We present a novel approach for investigating the spin structure of  hadrons based on the two-point function in quantum field theory. In a rotating frame, we derive two independent expressions of the two-point function and identify their equivalence,  which allows for a complete decomposition of the total spin of a composite system into the angular momenta of its constituent particles. Applying this approach to the proton, we analyze its spin structure in the massless quark limit. Our results indicate that the quark spin contribution accounts for approximately 27$\%$ of the total proton spin at a low-energy scale, significantly deviating from the prediction of the non-relativistic quark model.
\end{abstract}

\maketitle
Understanding the proton spin structure  remains a central challenge in hadron physics. The naive quark model initially predicted that its spin originates entirely from  three valence quarks.  However, the EMC experiment in the late 1980s \cite{EuropeanMuon:1987isl} revealed that quarks contribute only a small fraction, leading to the “proton spin puzzle”. Subsequent
experiments confirmed that quark spin accounts for about 30$\%$ of the total proton spin \cite{COMPASS:2006mhr,HERMES:2006jyl},
implying the significant roles of quark orbital angular momentum and gluons.  While various experimental studies, lattice QCD simulations, and theoretical approaches have provided valuable insights, a precise understanding of the spin structure has yet to be achieved (see review papers \cite{Deur:2018roz,Filippone:2001ux,Aidala:2012mv,Kuhn:2008sy}). As the upcoming Electron-Ion Collider (EIC) aims to resolve this issue \cite{Accardi:2012qut,AbdulKhalek:2021gbh}, the  spin structure of the proton is gaining more global attention.
In this work, we present a novel approach based on the two-point function in quantum field theory, which has not been widely utilized in conventional spin studies. By analyzing the two-point function in a rotating frame, we establish a  framework for hadron spin decomposition  and apply it to the proton in the massless quark limit.

In quantum field theory, the two-point function is the most elementary object for studying physical states associated with a quantum field $\phi(x)$:
\begin{align}
    \Pi(q)=i\int d^4 x e^{iqx}\langle 0|T\{\phi(x) \bar{\phi}(0)\}|0\rangle
    \label{twopoint}
\end{align}
If $\phi(x)$ is constructed from quark and/or gluon fields in quantum chromodynamics (QCD), it can create a hadronic state with the same quantum number as $\phi(x)$. However, since $\phi(x)$ is not an elementary field, it also creates excited  and multi-particle states. Thus, the two-point function typically includes the full spectrum of physical states sharing the same quantum number. Extracting a specific hadron, usually the ground state, from this spectrum is challenging due to the non-perturbative nature of QCD at low-energy scales where hadrons form. To address this, QCD Sum Rules (QCDSR) were developed by Shifman, Vainshtein, and Zakharov in 1979 \cite{Shifman:1978bx} and has since been widely used in hadron physics. This method is based on a dispersion relation that connects the two-point function in the physical $q^2$ region to that in the deep Euclidean region, $Q^2=-q^2\gg 0$, where the operator product expansion (OPE) is applicable.
In this framework, spectral properties of hadrons are constrained by a series of Wilson coefficients multiplied by QCD condensates, which are the vacuum expectation values of local QCD operators. Despite certain limitations, which will be discussed later, the QCDSR method has proven highly effective in elucidating hadron properties, particularly their masses, through QCD condensates that characterize non-perturbative dynamics (see review papers \cite{Reinders:1984sr,Colangelo:2000dp,Gubler:2018ctz}).

Thus far, most QCDSR studies have focused on analyzing the two-point function in an inertial frame to investigate hadron masses. While hadron spin information is also encoded in the two-point function, it is not readily apparent in an inertial frame. To reveal this, we consider the two-point function in a rotating frame with a uniform angular velocity $\Omega$ along the z-axis. Expanding  in $\Omega$, we define $\Pi^\rot(q)$ as the term proportional to $\Omega$, which captures the leading rotational corrections—a key quantity in this work. As we will demonstrate below, this first-order rotational term can be derived using two distinct methods, referred to as  Method A and  B  in this work.

In Method A, we directly evaluate Feynman diagrams that appear in Eq.\eqref{twopoint} in a rotating frame and extract terms proportional to $\Omega$.
In this frame, quark and gluon fields introduce spin and orbital angular momentum operators, as their rotational generators, at linear order in $\Omega$. Consequently, the sum of leading rotational corrections can be obtained by inserting the total angular momentum operator of elementary fields into the two-point function:
\begin{align}
\Pi_{\textrm{A}}^{\text{rot}}(q) = i \int d^4x \, d^4y \, e^{iqx} \langle 0 | T\{\phi(x)M_z(y)\bar{\phi}(0) \} | 0 \rangle, \label{methoda}
\end{align}
where $M_z=x^1T^{02}-x^2T^{01}$ represents the angular momentum density in the $z$-direction, with $T^{\mu\nu}$ being the symmetric energy-momentum tensor. 
In QCD, the total angular momentum density can be expressed as the sum of three gauge-invariant components \cite{Ji:1996ek}:
\begin{align}
\hspace*{-0.06cm}
\vec{M}(x)=\overbracket[0.06em][0.39em]{\frac{1}{2} \bar{\psi} \vec{\gamma} \gamma^5 \psi}^{\displaystyle{S_q}} + \overbracket[0.06em][0.73em]{\psi^\dagger (\vec{x} \times (-i\vec{D})) \psi}^{\displaystyle{L_q}}+ \overbracket[0.06em][0.73em]{\vec{x} \times (\vec{E} \times \vec{B})}^{\displaystyle{J_g}}, \label{jqcd}
\end{align}
where each term represents the contribution from quark spin $S_q$, quark orbital angular momentum $L_q$, and gluon total angular momentum $J_g$, respectively.
Here,  $\psi$ is the quark field, $\vec{D}=\vec{\partial}+ig\vec{A}$ is the covariant derivative, and $\vec{E}$ and $\vec{B}$ are the color electric and magnetic fields, respectively.

The evaluation of Eq.\eqref{methoda} is tedious yet relatively straightforward. The operators from Eq.\eqref{jqcd} are directly inserted into Feynman diagrams in an inertial frame, without requiring additional diagram types. The first two terms can be readily applied to the quark propagator, while the last term can be rewritten as a quark bilinear form using the equation of motion \cite{Balitsky:1997rs}. The result of this evaluation quantifies how quarks and gluons contribute to the total spin carried in the diagram. Thus, Eq.\eqref{methoda} plays a pivotal role in elucidating the spin structure of hadrons. 
However, it typically exhibits a more complex tensor and functional structure, as well as a more intricate spectral representation than in an inertial frame.
These complexities make it unclear what the best approach is for extracting the ground state from Eq.\eqref{methoda}.  For instance, Balitsky and Ji partially studied the gluon contribution to the proton spin using Eq.\eqref{methoda} in Ref.\cite{Balitsky:1997rs}, but their analysis focused on a single tensor structure.

Method B, on the other hand, offers a significantly simpler description, avoiding complex computations. 
In this method, we consider how the entire system transforms in the rotating frame, following the 
Lorentz transformation of the $\phi(x)$ field with a rotational angle of $-\Omega x^0$.  Substituting the transformed $\phi(x)$ field into Eq.\eqref{twopoint}, we find that, at linear order in $\Omega$, the total rotational effect on the two-point function takes the following simple form:
\begin{align}
    \Pi^{\rot}_{\textrm{B}}(q)=\big[S_z+ (\vec{q}\times(i\vec{\partial}_q))_3\big]\partial_q^0\Pi(q),
    \label{methodb}
\end{align}
where $S_z$ is the rotational generator (total spin operator) corresponding to the spin of $\phi(x)$ field and the second term accounts for the total orbital motion induced by the rotation. This expression shows that the first-order rotational term can be directly obtained from the two-point function in an inertial frame, without requiring any knowledge of the system’s internal structure.

So far, we have discussed two distinct methods.  Since both methods describe the same phenomenon, they must be equivalent:
$\Pi^{\rot}_{\textrm{A}}=\Pi^{\rot}_{\textrm{B}}$. 
To isolate the spin part, we assume the system is located at the center of rotation. Under this condition, the equivalence simplifies to:
\begin{align}
    \Pi^{\rot}_{\textrm{A}}(\omega)=S_z\frac{\partial \Pi(\omega)}{\partial\omega},
    \label{equivalence}
\end{align}
where $\omega$ denotes the system’s total energy.
This relation is a natural consequence of angular momentum conservation, yet it offers considerable practical advantages.
Since Eq.\eqref{equivalence} holds for each diagram individually, it allows us to discern how the total spin of $\phi$ field is distributed among the angular momenta of elementary particles in a given diagram.
Simultaneously, Eq.\eqref{equivalence} serves as a rigorous criterion for verifying computations of Eq.\eqref{methoda}. Any violation in a specific diagram suggests a potential error or missing term. Moreover, the intricate spectral representation in the rotating frame is fully determined by its counterpart in an inertial frame. 
Thus, this equivalence provides more useful insights for investigating hadron spin structure than relying solely on Eq.\eqref{methoda}.

This equivalence was first observed in the spin-1 case in a previous study by one  of the present authors \cite{Kim:2022vtt}. In that work, a spin-1 specific form of Eq.\eqref{equivalence} was derived using spin-rotation coupling  and spin structures of spin-1 heavy quarkonium states were analyzed. While the bottomonium results are comparable to the quark model, the charmonium results show a slightly smaller quark spin contribution. A similar tendency is also observed in a recent lattice simulation for charmed hadrons \cite{He:2024irz}.
These studies indicate that while the naive non-relativistic quark model remains useful for heavy quark systems, relativistic effects may suppress the quark spin contribution as the quark mass decreases. 

\begin{figure}
\centering
\begin{subfigure}{.15\textwidth}
  \centering
  \includegraphics[width=.9\linewidth]{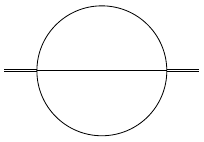}
  \caption{}
\end{subfigure}
\begin{subfigure}{.15\textwidth}
  \centering
  \includegraphics[width=.9\linewidth]{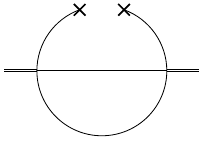}
  \caption{}
\end{subfigure}
\begin{subfigure}{.15\textwidth}
  \centering
  \includegraphics[width=.9\linewidth]{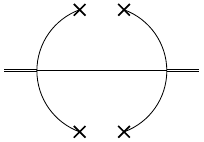}
  \caption{}
\end{subfigure}\\
\begin{subfigure}{.15\textwidth}
  \centering
  \includegraphics[width=.9\linewidth]{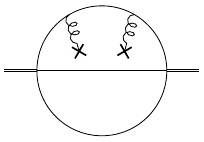}
  \caption{}
\end{subfigure}
\begin{subfigure}{.15\textwidth}
  \centering
  \includegraphics[width=.9\linewidth]{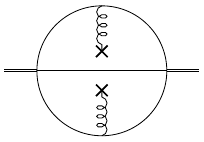}
  \caption{}
\end{subfigure}
\caption{Feynman diagrams for the proton}
\label{fig:diagram}
\end{figure}
In this work, we utilize this equivalence to explore the spin structure of the spin-1/2 proton in the massless quark limit. To this end, we first analyze the OPE of the two-point function for a composite field that couples strongly to the proton  \cite{Leinweber:1994nm,Yoo:2021gql}:
\begin{align}
    \phi(x)=\epsilon_{abc}[u^{a\rm{T}}(x)C\gamma^5d^b(x)]u^c(x),
\end{align}
where $a$, $b$, $c$ are color indices, $C$ is the charge conjugation matrix, $u$ and $d$ represent the up- and down-quark fields, respectively. For simplicity, we restrict our OPE series to the five diagrams in Fig.~\ref{fig:diagram}. Diagram (a) depicts the leading perturbative contribution from three free quarks. 
Diagrams (b-e) depict non-perturbative contributions, arising from two primary QCD condensates. The disconnected quark lines in (b-c) represent the quark condensate: $\VEV{\bar{q}q}=\VEV{0|\bar{u}u|0}=\VEV{0|\bar{d}d|0}$, while the disconnected curly lines in (d-e) represent the gluon condensate: $\VEV{G_0}=\VEV{0|\frac{\alpha_s}{\pi}G_{\mu\nu}G^{\mu\nu}|0}$.

The results of these diagrams are well known in an inertial frame (e.g., Ref.\cite{Cohen:1994wm}). After projecting them onto the positive-energy and spin-up state, we have:
\begin{align}
    \bar{\Pi}(\omega)/\omega&=-\frac{5\omega^4\log(-\omega^2)}{512\pi^4}+\frac{7\omega \log(-\omega^2)}{16\pi^2}\VEV{\bar{q}q}\nonumber\\
    &-\frac{7}{6\omega^2}\VEV{\bar{q}q}^2-\frac{5\log(-\omega^2)}{256\pi^2}\VEV{G_0}, \label{protonvac}
\end{align}
where  $\bar{\Pi}(\omega)\equiv\Tr[\Pi(\omega)\frac{1+\gamma^0}{2}\frac{1+\gamma^3\gamma^5}{2}]$   and polynomial terms in $\omega$ are omitted as they are irrelevant in this work. 
Applying Method B to Eq.\eqref{protonvac} readily yields the corresponding results in the rotating frame:
\begin{align}
        \bar{\Pi}_{\textrm{B}}^{\rm{rot}}(\omega)&=s_z    \frac{\partial\bar{\Pi}(\omega)}{\partial\omega}\nonumber\\
        &=-\frac{25\omega^4\log(-\omega^2)}{1024\pi^4}+\frac{7\omega\log(-\omega^2)}{16\pi^2}\VEV{\bar{q}q}\nonumber\\
        &+\frac{7}{12\omega^2}\VEV{\bar{q}q}^2-\frac{5\log(-\omega^2)}{512\pi^2}\VEV{G_0}, \label{protonrot}
\end{align}
where $s_z=1/2$ indicates the eigenvalue of $S_z$ for the spin-up state. By evaluating Eq.\eqref{methoda} for the same diagrams, we obtain $\bar{\Pi}_{\textrm{A}}^{\rm{rot}}(\omega)$ and confirm that the sum of all terms exactly reproduces Eq.\eqref{protonrot}. In each diagram, the total spin-1/2 is decomposed into contributions from quark spin, quark orbital, and gluon total angular momentum. Additionally, quark flavors are further distinguished as up-quark and down-quark. For quark fields forming the quark condensate, their transformations are treated separately and incorporated into Eq.\eqref{methoda}. The detailed spin decomposition results, with the sum of factors in parentheses normalized to 1, are given as follows:
\begin{widetext}
\begin{align}
    &&\langle S_u \rangle\quad\;\, &:&\langle S_d \rangle \quad\;\,&:&\langle L_u \rangle\quad\;\,&:&\langle L_d \rangle \quad \;\,&:&\langle J_g\rangle \quad\;\,&  \nonumber\\
    \text{diagram (a)} &:\; \Big(&-\frac{6}{25}\quad\;\;\, & + &-\frac{1}{25}\quad\;\;\,&+&\frac{36}{25}\quad\;\;\,&+&-\frac{4}{25}\quad\;\;\,&+&0\quad\;\;\;\,&\Big)\,\displaystyle\times\bigg[\frac{-25\omega^4\log(-\omega^2)}{1024\pi^4}\bigg],\\
  \text{diagram (b)} &:\; \Big(&\frac{6}{7} \quad\;\;\,& + &-\frac{1}{21}\quad\;\;\,&+&\frac{8}{21}\quad\;\;\,&+&-\frac{4}{21}\quad\;\;\,&+&0\quad\;\;\;\,&\Big)\,\displaystyle\times\bigg[\frac{7\omega\log(-\omega^2)}{16\pi^2}\VEV{\bar{q}q}\bigg],\\
    \text{diagram (c)} &:\; \Big(&\frac{10}{7}\quad\;\;\, & + &-\frac{3}{7}\quad\;\;\,&+&0\quad\;\;\;\,&+&0\quad\;\;\;\,&+&0\quad\;\;\;\,&\Big)\,\displaystyle\times\bigg[\frac{7}{12\omega^2}\VEV{\bar{q}q}^2\bigg],\\
    \text{diagram (d-e)}&:\; \Big(&-\frac{2}{5} \quad\;\;\,& + &-\frac{11}{15}\quad\;\;\,&+&\frac{40}{9}\quad\;\;\,&+&\frac{56}{45}\quad\;\;\,&+&-\frac{32}{9}\quad\;\;\,&\Big)\,\displaystyle\times\bigg[\frac{-5\log(-\omega^2)}{512\pi^2}\VEV{G_0}\bigg].
\end{align}
\end{widetext} 

Meanwhile, the spectral representation in an inertial frame reveals that $\bar{\Pi}^\rot_\textrm{B}$ has a ground state corresponding to the spin-up proton, with a residue proportional to $s_z$. In the QCDSR method, this ground state contribution can be extracted through a sum rule:
\begin{align}
\hspace*{-0.094cm}
\int^{s_0}_0ds \,\textrm{Im}\bar{\Pi}^\rot_{\textrm{B}}(s)e^{-\frac{s}{M^2}}=2\pi s_z\lambda_\textrm{p}^2\Big(\frac{2m^2_\textrm{p}}{M^2}-1\Big)e^{-\frac{m_\textrm{p}^2}{M^2}}, \label{sumruleb}
\end{align}
where $\lambda_\textrm{p}$ is the coupling strength between the proton and the $\phi$ field, $m_\textrm{p}$ is the proton mass, $M$ is an unphysical parameter called the Borel mass, and $s_0$ is a threshold parameter that suppresses continuum contributions.
Since the left-hand side of Eq.\eqref{sumruleb} can be decomposed into terms based on their spin origins, the spin structure  can be revealed  from the following ratio:
\begin{align}
    \frac{\displaystyle\int^{s_0}_0ds \,\textrm{Im}\bar{\Pi}^\rot_{\textrm{A}}(s)e^{-\frac{s}{M^2}}}{\displaystyle\int^{s_0}_0ds\, \textrm{Im}\bar{\Pi}^\rot_{\textrm{B}}(s)e^{-\frac{s}{M^2}}}=\frac{s_q+l_q+j_g}{s_z}=1, \label{spinstructure}
\end{align}
where spin-independent terms cancel out, and $s_q$, $l_q$, and $j_g$ denote the spin components of the proton at rest,  corresponding to quark spin, quark orbital, and gluon total angular momentum, respectively. While this ratio always equals 1 by definition, the individual components vary with the Borel mass $M$ and the threshold parameter $s_0$. Since $M$ is an auxiliary parameter, a reliable sum rule should exhibit minimal sensitivity to it, with $s_0$ chosen to reduce this dependence.
\begin{figure}
    \includegraphics[width=\linewidth]{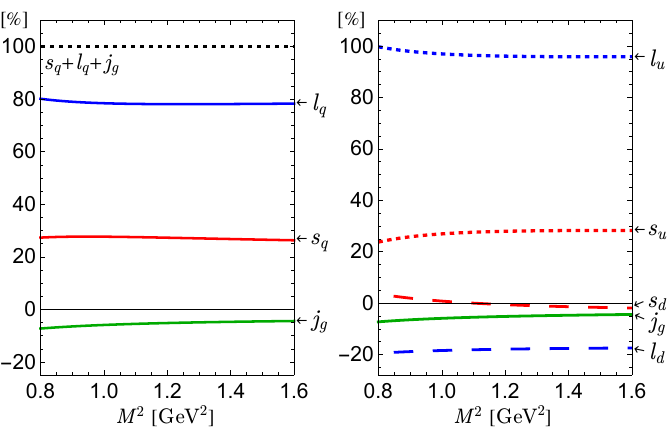}
    \caption{\justifying The left panel decomposes the proton spin into $s_q$, $l_q$, and $j_g$, with the dashed black line representing $s_q+l_q+j_g=s_z$. The right panel further resolves $s_q$ and $l_q$ into up- and down-quark contributions.}
    \label{fig:plot}
\end{figure}

Fig.~\ref{fig:plot} shows the ratios of the individual spin components to the total proton spin as functions of $M^2$. 
For numerical analysis, we use the condensate values $\langle \bar{q}q \rangle = -(0.24 \, \rm{GeV})^3$ \cite{Ioffe:1981kw,Ioffe:1983ju} and $\langle G_0 \rangle = 0.012 \, \rm{GeV}^4$ \cite{Shifman:1978bx}, while setting $s_0 = 3 \, \rm{GeV}^2$.
Practically, the sum rule is not reliable over the entire $M$ range. For $M^2<0.8\,\textrm{GeV}^2$, the $\VEV{\bar{q}q}^2$ term exceeds 20\% of the total OPE, indicating poor convergence of the truncated series.  Conversely, for $M^2>1.6\,\textrm{GeV}^2$, the continuum contribution exceeds 60\% of the total, potentially contaminating the ground-state signal.  To ensure reliability, we restrict our sum rule to $M^2 \in [0.8, 1.6]\,\mathrm{GeV}^2$ and take averages over this range to read off the spin component values. The final results are summarized in TABLE~\ref{tab:spin_decomposition}, with uncertainties given as standard deviations.
\begin{table}[H]
  \centering
  \caption{Spin decomposition of the proton at rest}
  \label{tab:spin_decomposition}
  \begin{tabular}{cccc}
    \toprule
    Component & Overall (\%) & Up-quark (\%) & Down-quark (\%) \\
    \midrule
    $s_q$ & $27.1\pm0.4$             & $27.3 \pm 1.1$   & $\;\,-0.2 \pm 1.5$ \\
    $l_q$ & $78.3\pm0.5$             & $96.4 \pm 0.6$   & $-18.2 \pm 0.6$ \\
    $j_g$ & $-5.4 \pm 0.8\,$ & ---              & --- \\
    \bottomrule
  \end{tabular}
\end{table}

Our result reveals a key qualitative feature: the quark spin contribution is significantly small, markedly deviating from the naive quark model's prediction. Here, we observe that the quark condensate terms, i.e., diagrams (b) and (c), contribute about 50$\%$ of the total proton spin, while the remaining 50$\%$ primarily comes from the perturbative diagram. As a direct indicator of spontaneous chiral symmetry breaking in QCD, the quark condensate has long been recognized as essential for understanding the origin of the proton mass. This study implies that it  significantly affects not only the proton mass but also its spin structure.

On the other hand, the results for other components appear to contradict recent lattice simulations \cite{LHPC:2010jcs,Gockeler:2003jfa,Alexandrou:2013joa,Deka:2013zha}, which predict $l_u<0$, $l_d>0$, and $j_g>0$. However, since spin components are scale-dependent quantities, identifying the renormalization scale $\mu$ is important. Unfortunately,  at the current OPE series, the scale (and scheme) dependence is poorly determined. In practice, it only remains in the quark condensate $\VEV{\bar{q}q}$, suggesting that our results are effectively evaluated at the scale where this condensate is determined. Based on recent lattice averaged values \cite{FlavourLatticeAveragingGroup:2019iem}, we infer that our $\VEV{\bar{q}q}$ value corresponds to a lower scale, approximately $\mu = 0.5-1 \, \mathrm{GeV}$, which directly leads to our scale. Since most lattice simulation results are provided at higher scales, typically $\mu\simeq2\,\mathrm{GeV}$ in the $\overline{\text{MS}}$ scheme, the effect of QCD evolution must be considered. Indeed, it has been shown that during LO evolution, the signs of $l_u$ and $l_d$ reverse below $\mu \approx 1\,\mathrm{GeV}$, while $s_q$ remains scale-independent \cite{Thomas:2008ga}. As the scale decreases, $l_u$ shifts from negative to large positive values, while $l_d$ transitions from positive to large negative values at slightly lower scales. Comparable trends were also observed in the NLO and NNLO studies of Ref.\cite{Altenbuchinger:2010sz}, where $j_g$ similarly shifts to negative values at low-energy scales. Thus, while specifying the scale value remains somewhat challenging at the present stage, our results appear to exhibit the qualitatively consistent features expected at low-energy scales.

To achieve more accurate predictions, several improvements in the OPE series are possible. 
First, it will be important to incorporate $\mathcal{O}(\alpha_s)$ perturbative corrections into the diagrams, 
which  were already considered in an inertial frame \cite{Ovchinnikov:1991mu}. In particular, the  correction to the leading perturbative diagram—responsible for one-gluon exchange between quarks—are crucial, as this process significantly influences the redistribution of spin components \cite{Myhrer:2007cf}. Moreover, since the explicit $\mu$ dependence of Wilson coefficients begins to  take effect at $\mathcal{O}(\alpha_s)$, these corrections are vital for capturing the scale dependence of the spin components. 
Second, one can systematically extend the OPE series by including higher-dimensional operators. While the condensates $\langle \bar{q}q \rangle$ and $\langle G_0 \rangle$ play an important role, they  may be insufficient to fully capture the non-perturbative dynamics. 
Including various condensates of higher dimension in an inertial frame  was already shown to improve the accuracy of QCDSR predictions for the nucleon mass \cite{Ioffe:1983ju}. Third, finite quark mass corrections deserve consideration, as they explicitly break chiral symmetry in QCD and  clarify the distinction between up- and down-quarks.
Finally, running effects from the coupling constant, quark masses, and anomalous dimensions of operators are also necessary for a complete understanding of the scale dependence. 

In summary, we present  a novel pragmatic approach to exploring the spin structure of hadrons based on the two-point function. In a rotating frame, we describe the two-point function using two methods: one explains the system’s rotation in terms of angular momenta of its constituent particles, while the other interprets it based on the total spin of the system. 
The equivalence between these methods offers valuable insight into the spin structure of the two-point function. 
Once the ground-state contribution is extracted via the QCDSR method, the ratio of these two methods effectively reveals the spin structure of the corresponding hadron.
In this work, we applied this framework to qualitatively explore the spin structure of the proton in the massless quark limit. While further and systematic improvements are possible, 
our results already demonstrate the important qualitative feature that the quark spin contribution is small. 
With an improved OPE, the two-point function will provide more  accurate information on the spin structure of the proton.

\begin{acknowledgments}
This work is supported by the WPI program “Sustainability with Knotted Chiral Meta Matter (WPI-SKCM$^2$)” at Hiroshima University.
The work of C.S. was supported partly by the Polish National Science Centre (NCN) under OPUS Grant No. 2022/45/B/ST2/01527. 
P.G. is supported by the Grant-in-Aid for Scientific Research (A) (JSPS KAKENHI Grant Number JP22H00122).
\end{acknowledgments}

\bibliography{apssamp}

\providecommand{\noopsort}[1]{}\providecommand{\singleletter}[1]{#1}%
\begin{thebibliography}{31}%
\makeatletter
\providecommand \@ifxundefined [1]{%
 \@ifx{#1\undefined}
}%
\providecommand \@ifnum [1]{%
 \ifnum #1\expandafter \@firstoftwo
 \else \expandafter \@secondoftwo
 \fi
}%
\providecommand \@ifx [1]{%
 \ifx #1\expandafter \@firstoftwo
 \else \expandafter \@secondoftwo
 \fi
}%
\providecommand \natexlab [1]{#1}%
\providecommand \enquote  [1]{``#1''}%
\providecommand \bibnamefont  [1]{#1}%
\providecommand \bibfnamefont [1]{#1}%
\providecommand \citenamefont [1]{#1}%
\providecommand \href@noop [0]{\@secondoftwo}%
\providecommand \href [0]{\begingroup \@sanitize@url \@href}%
\providecommand \@href[1]{\@@startlink{#1}\@@href}%
\providecommand \@@href[1]{\endgroup#1\@@endlink}%
\providecommand \@sanitize@url [0]{\catcode `\\12\catcode `\$12\catcode `\&12\catcode `\#12\catcode `\^12\catcode `\_12\catcode `\%12\relax}%
\providecommand \@@startlink[1]{}%
\providecommand \@@endlink[0]{}%
\providecommand \url  [0]{\begingroup\@sanitize@url \@url }%
\providecommand \@url [1]{\endgroup\@href {#1}{\urlprefix }}%
\providecommand \urlprefix  [0]{URL }%
\providecommand \Eprint [0]{\href }%
\providecommand \doibase [0]{https://doi.org/}%
\providecommand \selectlanguage [0]{\@gobble}%
\providecommand \bibinfo  [0]{\@secondoftwo}%
\providecommand \bibfield  [0]{\@secondoftwo}%
\providecommand \translation [1]{[#1]}%
\providecommand \BibitemOpen [0]{}%
\providecommand \bibitemStop [0]{}%
\providecommand \bibitemNoStop [0]{.\EOS\space}%
\providecommand \EOS [0]{\spacefactor3000\relax}%
\providecommand \BibitemShut  [1]{\csname bibitem#1\endcsname}%
\let\auto@bib@innerbib\@empty
\bibitem [{\citenamefont {Ashman}\ \emph {et~al.}(1988)\citenamefont {Ashman} \emph {et~al.}}]{EuropeanMuon:1987isl}%
  \BibitemOpen
  \bibfield  {author} {\bibinfo {author} {\bibfnamefont {J.}~\bibnamefont {Ashman}} \emph {et~al.} (\bibinfo {collaboration} {European Muon}),\ }\bibfield  {title} {\bibinfo {title} {{A Measurement of the Spin Asymmetry and Determination of the Structure Function g(1) in Deep Inelastic Muon-Proton Scattering}},\ }\href {https://doi.org/10.1016/0370-2693(88)91523-7} {\bibfield  {journal} {\bibinfo  {journal} {Phys. Lett. B}\ }\textbf {\bibinfo {volume} {206}},\ \bibinfo {pages} {364} (\bibinfo {year} {1988})}\BibitemShut {NoStop}%
\bibitem [{\citenamefont {Alexakhin}\ \emph {et~al.}(2007)\citenamefont {Alexakhin} \emph {et~al.}}]{COMPASS:2006mhr}%
  \BibitemOpen
  \bibfield  {author} {\bibinfo {author} {\bibfnamefont {V.~Y.}\ \bibnamefont {Alexakhin}} \emph {et~al.} (\bibinfo {collaboration} {COMPASS}),\ }\bibfield  {title} {\bibinfo {title} {{The Deuteron Spin-dependent Structure Function g1(d) and its First Moment}},\ }\href {https://doi.org/10.1016/j.physletb.2006.12.076} {\bibfield  {journal} {\bibinfo  {journal} {Phys. Lett. B}\ }\textbf {\bibinfo {volume} {647}},\ \bibinfo {pages} {8} (\bibinfo {year} {2007})},\ \Eprint {https://arxiv.org/abs/hep-ex/0609038} {arXiv:hep-ex/0609038} \BibitemShut {NoStop}%
\bibitem [{\citenamefont {Airapetian}\ \emph {et~al.}(2007)\citenamefont {Airapetian} \emph {et~al.}}]{HERMES:2006jyl}%
  \BibitemOpen
  \bibfield  {author} {\bibinfo {author} {\bibfnamefont {A.}~\bibnamefont {Airapetian}} \emph {et~al.} (\bibinfo {collaboration} {HERMES}),\ }\bibfield  {title} {\bibinfo {title} {{Precise determination of the spin structure function g(1) of the proton, deuteron and neutron}},\ }\href {https://doi.org/10.1103/PhysRevD.75.012007} {\bibfield  {journal} {\bibinfo  {journal} {Phys. Rev. D}\ }\textbf {\bibinfo {volume} {75}},\ \bibinfo {pages} {012007} (\bibinfo {year} {2007})},\ \Eprint {https://arxiv.org/abs/hep-ex/0609039} {arXiv:hep-ex/0609039} \BibitemShut {NoStop}%
\bibitem [{\citenamefont {Deur}\ \emph {et~al.}(2019)\citenamefont {Deur}, \citenamefont {Brodsky},\ and\ \citenamefont {De~T\'eramond}}]{Deur:2018roz}%
  \BibitemOpen
  \bibfield  {author} {\bibinfo {author} {\bibfnamefont {A.}~\bibnamefont {Deur}}, \bibinfo {author} {\bibfnamefont {S.~J.}\ \bibnamefont {Brodsky}},\ and\ \bibinfo {author} {\bibfnamefont {G.~F.}\ \bibnamefont {De~T\'eramond}},\ }\bibfield  {title} {\bibinfo {title} {{The Spin Structure of the Nucleon}},\ }\href {https://doi.org/10.1088/1361-6633/ab0b8f} {\bibfield  {journal} {\bibinfo  {journal} {Rept. Prog. Phys.}\ }\textbf {\bibinfo {volume} {82}},\ \bibinfo {pages} {076201} (\bibinfo {year} {2019})},\ \Eprint {https://arxiv.org/abs/1807.05250} {arXiv:1807.05250 [hep-ph]} \BibitemShut {NoStop}%
\bibitem [{\citenamefont {Filippone}\ and\ \citenamefont {Ji}(2001)}]{Filippone:2001ux}%
  \BibitemOpen
  \bibfield  {author} {\bibinfo {author} {\bibfnamefont {B.~W.}\ \bibnamefont {Filippone}}\ and\ \bibinfo {author} {\bibfnamefont {X.-D.}\ \bibnamefont {Ji}},\ }\bibfield  {title} {\bibinfo {title} {{The Spin structure of the nucleon}},\ }\href {https://doi.org/10.1007/0-306-47915-X_1} {\bibfield  {journal} {\bibinfo  {journal} {Adv. Nucl. Phys.}\ }\textbf {\bibinfo {volume} {26}},\ \bibinfo {pages} {1} (\bibinfo {year} {2001})},\ \Eprint {https://arxiv.org/abs/hep-ph/0101224} {arXiv:hep-ph/0101224} \BibitemShut {NoStop}%
\bibitem [{\citenamefont {Aidala}\ \emph {et~al.}(2013)\citenamefont {Aidala}, \citenamefont {Bass}, \citenamefont {Hasch},\ and\ \citenamefont {Mallot}}]{Aidala:2012mv}%
  \BibitemOpen
  \bibfield  {author} {\bibinfo {author} {\bibfnamefont {C.~A.}\ \bibnamefont {Aidala}}, \bibinfo {author} {\bibfnamefont {S.~D.}\ \bibnamefont {Bass}}, \bibinfo {author} {\bibfnamefont {D.}~\bibnamefont {Hasch}},\ and\ \bibinfo {author} {\bibfnamefont {G.~K.}\ \bibnamefont {Mallot}},\ }\bibfield  {title} {\bibinfo {title} {{The Spin Structure of the Nucleon}},\ }\href {https://doi.org/10.1103/RevModPhys.85.655} {\bibfield  {journal} {\bibinfo  {journal} {Rev. Mod. Phys.}\ }\textbf {\bibinfo {volume} {85}},\ \bibinfo {pages} {655} (\bibinfo {year} {2013})},\ \Eprint {https://arxiv.org/abs/1209.2803} {arXiv:1209.2803 [hep-ph]} \BibitemShut {NoStop}%
\bibitem [{\citenamefont {Kuhn}\ \emph {et~al.}(2009)\citenamefont {Kuhn}, \citenamefont {Chen},\ and\ \citenamefont {Leader}}]{Kuhn:2008sy}%
  \BibitemOpen
  \bibfield  {author} {\bibinfo {author} {\bibfnamefont {S.~E.}\ \bibnamefont {Kuhn}}, \bibinfo {author} {\bibfnamefont {J.~P.}\ \bibnamefont {Chen}},\ and\ \bibinfo {author} {\bibfnamefont {E.}~\bibnamefont {Leader}},\ }\bibfield  {title} {\bibinfo {title} {{Spin Structure of the Nucleon - Status and Recent Results}},\ }\href {https://doi.org/10.1016/j.ppnp.2009.02.001} {\bibfield  {journal} {\bibinfo  {journal} {Prog. Part. Nucl. Phys.}\ }\textbf {\bibinfo {volume} {63}},\ \bibinfo {pages} {1} (\bibinfo {year} {2009})},\ \Eprint {https://arxiv.org/abs/0812.3535} {arXiv:0812.3535 [hep-ph]} \BibitemShut {NoStop}%
\bibitem [{\citenamefont {Accardi}\ \emph {et~al.}(2016)\citenamefont {Accardi} \emph {et~al.}}]{Accardi:2012qut}%
  \BibitemOpen
  \bibfield  {author} {\bibinfo {author} {\bibfnamefont {A.}~\bibnamefont {Accardi}} \emph {et~al.},\ }\bibfield  {title} {\bibinfo {title} {{Electron Ion Collider: The Next QCD Frontier}: {Understanding the glue that binds us all}},\ }\href {https://doi.org/10.1140/epja/i2016-16268-9} {\bibfield  {journal} {\bibinfo  {journal} {Eur. Phys. J. A}\ }\textbf {\bibinfo {volume} {52}},\ \bibinfo {pages} {268} (\bibinfo {year} {2016})},\ \Eprint {https://arxiv.org/abs/1212.1701} {arXiv:1212.1701 [nucl-ex]} \BibitemShut {NoStop}%
\bibitem [{\citenamefont {Abdul~Khalek}\ \emph {et~al.}(2022)\citenamefont {Abdul~Khalek} \emph {et~al.}}]{AbdulKhalek:2021gbh}%
  \BibitemOpen
  \bibfield  {author} {\bibinfo {author} {\bibfnamefont {R.}~\bibnamefont {Abdul~Khalek}} \emph {et~al.},\ }\bibfield  {title} {\bibinfo {title} {{Science Requirements and Detector Concepts for the Electron-Ion Collider}: {EIC Yellow Report}},\ }\href {https://doi.org/10.1016/j.nuclphysa.2022.122447} {\bibfield  {journal} {\bibinfo  {journal} {Nucl. Phys. A}\ }\textbf {\bibinfo {volume} {1026}},\ \bibinfo {pages} {122447} (\bibinfo {year} {2022})},\ \Eprint {https://arxiv.org/abs/2103.05419} {arXiv:2103.05419 [physics.ins-det]} \BibitemShut {NoStop}%
\bibitem [{\citenamefont {Shifman}\ \emph {et~al.}(1979)\citenamefont {Shifman}, \citenamefont {Vainshtein},\ and\ \citenamefont {Zakharov}}]{Shifman:1978bx}%
  \BibitemOpen
  \bibfield  {author} {\bibinfo {author} {\bibfnamefont {M.~A.}\ \bibnamefont {Shifman}}, \bibinfo {author} {\bibfnamefont {A.~I.}\ \bibnamefont {Vainshtein}},\ and\ \bibinfo {author} {\bibfnamefont {V.~I.}\ \bibnamefont {Zakharov}},\ }\bibfield  {title} {\bibinfo {title} {{QCD and Resonance Physics. Theoretical Foundations}},\ }\href {https://doi.org/10.1016/0550-3213(79)90022-1} {\bibfield  {journal} {\bibinfo  {journal} {Nucl. Phys. B}\ }\textbf {\bibinfo {volume} {147}},\ \bibinfo {pages} {385} (\bibinfo {year} {1979})}\BibitemShut {NoStop}%
\bibitem [{\citenamefont {Reinders}\ \emph {et~al.}(1985)\citenamefont {Reinders}, \citenamefont {Rubinstein},\ and\ \citenamefont {Yazaki}}]{Reinders:1984sr}%
  \BibitemOpen
  \bibfield  {author} {\bibinfo {author} {\bibfnamefont {L.~J.}\ \bibnamefont {Reinders}}, \bibinfo {author} {\bibfnamefont {H.}~\bibnamefont {Rubinstein}},\ and\ \bibinfo {author} {\bibfnamefont {S.}~\bibnamefont {Yazaki}},\ }\bibfield  {title} {\bibinfo {title} {{Hadron Properties from QCD Sum Rules}},\ }\href {https://doi.org/10.1016/0370-1573(85)90065-1} {\bibfield  {journal} {\bibinfo  {journal} {Phys. Rept.}\ }\textbf {\bibinfo {volume} {127}},\ \bibinfo {pages} {1} (\bibinfo {year} {1985})}\BibitemShut {NoStop}%
\bibitem [{\citenamefont {Colangelo}\ and\ \citenamefont {Khodjamirian}(2000)}]{Colangelo:2000dp}%
  \BibitemOpen
  \bibfield  {author} {\bibinfo {author} {\bibfnamefont {P.}~\bibnamefont {Colangelo}}\ and\ \bibinfo {author} {\bibfnamefont {A.}~\bibnamefont {Khodjamirian}},\ }\href {https://doi.org/10.1142/9789812810458_0033} {\bibinfo {title} {{QCD sum rules, a modern perspective}}} (\bibinfo {year} {2000}),\ \Eprint {https://arxiv.org/abs/hep-ph/0010175} {arXiv:hep-ph/0010175} \BibitemShut {NoStop}%
\bibitem [{\citenamefont {Gubler}\ and\ \citenamefont {Satow}(2019)}]{Gubler:2018ctz}%
  \BibitemOpen
  \bibfield  {author} {\bibinfo {author} {\bibfnamefont {P.}~\bibnamefont {Gubler}}\ and\ \bibinfo {author} {\bibfnamefont {D.}~\bibnamefont {Satow}},\ }\bibfield  {title} {\bibinfo {title} {{Recent Progress in QCD Condensate Evaluations and Sum Rules}},\ }\href {https://doi.org/10.1016/j.ppnp.2019.02.005} {\bibfield  {journal} {\bibinfo  {journal} {Prog. Part. Nucl. Phys.}\ }\textbf {\bibinfo {volume} {106}},\ \bibinfo {pages} {1} (\bibinfo {year} {2019})},\ \Eprint {https://arxiv.org/abs/1812.00385} {arXiv:1812.00385 [hep-ph]} \BibitemShut {NoStop}%
\bibitem [{\citenamefont {Ji}(1997)}]{Ji:1996ek}%
  \BibitemOpen
  \bibfield  {author} {\bibinfo {author} {\bibfnamefont {X.-D.}\ \bibnamefont {Ji}},\ }\bibfield  {title} {\bibinfo {title} {{Gauge-Invariant Decomposition of Nucleon Spin}},\ }\href {https://doi.org/10.1103/PhysRevLett.78.610} {\bibfield  {journal} {\bibinfo  {journal} {Phys. Rev. Lett.}\ }\textbf {\bibinfo {volume} {78}},\ \bibinfo {pages} {610} (\bibinfo {year} {1997})},\ \Eprint {https://arxiv.org/abs/hep-ph/9603249} {arXiv:hep-ph/9603249} \BibitemShut {NoStop}%
\bibitem [{\citenamefont {Balitsky}\ and\ \citenamefont {Ji}(1997)}]{Balitsky:1997rs}%
  \BibitemOpen
  \bibfield  {author} {\bibinfo {author} {\bibfnamefont {I.}~\bibnamefont {Balitsky}}\ and\ \bibinfo {author} {\bibfnamefont {X.-D.}\ \bibnamefont {Ji}},\ }\bibfield  {title} {\bibinfo {title} {{How much of the nucleon spin is carried by glue?}},\ }\href {https://doi.org/10.1103/PhysRevLett.79.1225} {\bibfield  {journal} {\bibinfo  {journal} {Phys. Rev. Lett.}\ }\textbf {\bibinfo {volume} {79}},\ \bibinfo {pages} {1225} (\bibinfo {year} {1997})},\ \Eprint {https://arxiv.org/abs/hep-ph/9702277} {arXiv:hep-ph/9702277} \BibitemShut {NoStop}%
\bibitem [{\citenamefont {Kim}\ \emph {et~al.}(2023)\citenamefont {Kim}, \citenamefont {Lee},\ and\ \citenamefont {Cho}}]{Kim:2022vtt}%
  \BibitemOpen
  \bibfield  {author} {\bibinfo {author} {\bibfnamefont {H.}~\bibnamefont {Kim}}, \bibinfo {author} {\bibfnamefont {S.~H.}\ \bibnamefont {Lee}},\ and\ \bibinfo {author} {\bibfnamefont {S.}~\bibnamefont {Cho}},\ }\bibfield  {title} {\bibinfo {title} {{Spin-1 quarkonia in a rotating frame and their spin contents}},\ }\href {https://doi.org/10.1016/j.physletb.2023.137986} {\bibfield  {journal} {\bibinfo  {journal} {Phys. Lett. B}\ }\textbf {\bibinfo {volume} {843}},\ \bibinfo {pages} {137986} (\bibinfo {year} {2023})},\ \Eprint {https://arxiv.org/abs/2212.14570} {arXiv:2212.14570 [hep-ph]} \BibitemShut {NoStop}%
\bibitem [{\citenamefont {He}\ \emph {et~al.}(2024)\citenamefont {He}, \citenamefont {Liang},\ and\ \citenamefont {Yang}}]{He:2024irz}%
  \BibitemOpen
  \bibfield  {author} {\bibinfo {author} {\bibfnamefont {F.}~\bibnamefont {He}}, \bibinfo {author} {\bibfnamefont {J.}~\bibnamefont {Liang}},\ and\ \bibinfo {author} {\bibfnamefont {Y.-B.}\ \bibnamefont {Yang}},\ }\href@noop {} {\bibinfo {title} {{Origin of hadron spin based on Lattice QCD study on the charmed hadrons}}} (\bibinfo {year} {2024}),\ \Eprint {https://arxiv.org/abs/2410.08046} {arXiv:2410.08046 [hep-lat]} \BibitemShut {NoStop}%
\bibitem [{\citenamefont {Leinweber}(1995)}]{Leinweber:1994nm}%
  \BibitemOpen
  \bibfield  {author} {\bibinfo {author} {\bibfnamefont {D.~B.}\ \bibnamefont {Leinweber}},\ }\bibfield  {title} {\bibinfo {title} {{Nucleon properties from unconventional interpolating fields}},\ }\href {https://doi.org/10.1103/PhysRevD.51.6383} {\bibfield  {journal} {\bibinfo  {journal} {Phys. Rev. D}\ }\textbf {\bibinfo {volume} {51}},\ \bibinfo {pages} {6383} (\bibinfo {year} {1995})},\ \Eprint {https://arxiv.org/abs/nucl-th/9406001} {arXiv:nucl-th/9406001} \BibitemShut {NoStop}%
\bibitem [{\citenamefont {Yoo}\ \emph {et~al.}(2022)\citenamefont {Yoo}, \citenamefont {Aoki}, \citenamefont {Boyle}, \citenamefont {Izubuchi}, \citenamefont {Soni},\ and\ \citenamefont {Syritsyn}}]{Yoo:2021gql}%
  \BibitemOpen
  \bibfield  {author} {\bibinfo {author} {\bibfnamefont {J.-S.}\ \bibnamefont {Yoo}}, \bibinfo {author} {\bibfnamefont {Y.}~\bibnamefont {Aoki}}, \bibinfo {author} {\bibfnamefont {P.}~\bibnamefont {Boyle}}, \bibinfo {author} {\bibfnamefont {T.}~\bibnamefont {Izubuchi}}, \bibinfo {author} {\bibfnamefont {A.}~\bibnamefont {Soni}},\ and\ \bibinfo {author} {\bibfnamefont {S.}~\bibnamefont {Syritsyn}},\ }\bibfield  {title} {\bibinfo {title} {{Proton decay matrix elements on the lattice at physical pion mass}},\ }\href {https://doi.org/10.1103/PhysRevD.105.074501} {\bibfield  {journal} {\bibinfo  {journal} {Phys. Rev. D}\ }\textbf {\bibinfo {volume} {105}},\ \bibinfo {pages} {074501} (\bibinfo {year} {2022})},\ \Eprint {https://arxiv.org/abs/2111.01608} {arXiv:2111.01608 [hep-lat]} \BibitemShut {NoStop}%
\bibitem [{\citenamefont {Cohen}\ \emph {et~al.}(1995)\citenamefont {Cohen}, \citenamefont {Furnstahl}, \citenamefont {Griegel},\ and\ \citenamefont {Jin}}]{Cohen:1994wm}%
  \BibitemOpen
  \bibfield  {author} {\bibinfo {author} {\bibfnamefont {T.~D.}\ \bibnamefont {Cohen}}, \bibinfo {author} {\bibfnamefont {R.~J.}\ \bibnamefont {Furnstahl}}, \bibinfo {author} {\bibfnamefont {D.~K.}\ \bibnamefont {Griegel}},\ and\ \bibinfo {author} {\bibfnamefont {X.-m.}\ \bibnamefont {Jin}},\ }\bibfield  {title} {\bibinfo {title} {{QCD sum rules and applications to nuclear physics}},\ }\href {https://doi.org/10.1016/0146-6410(95)00043-I} {\bibfield  {journal} {\bibinfo  {journal} {Prog. Part. Nucl. Phys.}\ }\textbf {\bibinfo {volume} {35}},\ \bibinfo {pages} {221} (\bibinfo {year} {1995})},\ \Eprint {https://arxiv.org/abs/hep-ph/9503315} {arXiv:hep-ph/9503315} \BibitemShut {NoStop}%
\bibitem [{\citenamefont {Ioffe}(1981)}]{Ioffe:1981kw}%
  \BibitemOpen
  \bibfield  {author} {\bibinfo {author} {\bibfnamefont {B.~L.}\ \bibnamefont {Ioffe}},\ }\bibfield  {title} {\bibinfo {title} {{Calculation of Baryon Masses in Quantum Chromodynamics}},\ }\href {https://doi.org/10.1016/0550-3213(81)90259-5} {\bibfield  {journal} {\bibinfo  {journal} {Nucl. Phys. B}\ }\textbf {\bibinfo {volume} {188}},\ \bibinfo {pages} {317} (\bibinfo {year} {1981})},\ \bibinfo {note} {[Erratum: Nucl.Phys.B 191, 591--592 (1981)]}\BibitemShut {NoStop}%
\bibitem [{\citenamefont {Ioffe}\ and\ \citenamefont {Smilga}(1984)}]{Ioffe:1983ju}%
  \BibitemOpen
  \bibfield  {author} {\bibinfo {author} {\bibfnamefont {B.~L.}\ \bibnamefont {Ioffe}}\ and\ \bibinfo {author} {\bibfnamefont {A.~V.}\ \bibnamefont {Smilga}},\ }\bibfield  {title} {\bibinfo {title} {{Nucleon Magnetic Moments and Magnetic Properties of Vacuum in QCD}},\ }\href {https://doi.org/10.1016/0550-3213(84)90364-X} {\bibfield  {journal} {\bibinfo  {journal} {Nucl. Phys. B}\ }\textbf {\bibinfo {volume} {232}},\ \bibinfo {pages} {109} (\bibinfo {year} {1984})}\BibitemShut {NoStop}%
\bibitem [{\citenamefont {Bratt}\ \emph {et~al.}(2010)\citenamefont {Bratt} \emph {et~al.}}]{LHPC:2010jcs}%
  \BibitemOpen
  \bibfield  {author} {\bibinfo {author} {\bibfnamefont {J.~D.}\ \bibnamefont {Bratt}} \emph {et~al.} (\bibinfo {collaboration} {LHPC}),\ }\bibfield  {title} {\bibinfo {title} {{Nucleon structure from mixed action calculations using 2+1 flavors of asqtad sea and domain wall valence fermions}},\ }\href {https://doi.org/10.1103/PhysRevD.82.094502} {\bibfield  {journal} {\bibinfo  {journal} {Phys. Rev. D}\ }\textbf {\bibinfo {volume} {82}},\ \bibinfo {pages} {094502} (\bibinfo {year} {2010})},\ \Eprint {https://arxiv.org/abs/1001.3620} {arXiv:1001.3620 [hep-lat]} \BibitemShut {NoStop}%
\bibitem [{\citenamefont {Gockeler}\ \emph {et~al.}(2004)\citenamefont {Gockeler}, \citenamefont {Horsley}, \citenamefont {Pleiter}, \citenamefont {Rakow}, \citenamefont {Schafer}, \citenamefont {Schierholz},\ and\ \citenamefont {Schroers}}]{Gockeler:2003jfa}%
  \BibitemOpen
  \bibfield  {author} {\bibinfo {author} {\bibfnamefont {M.}~\bibnamefont {Gockeler}}, \bibinfo {author} {\bibfnamefont {R.}~\bibnamefont {Horsley}}, \bibinfo {author} {\bibfnamefont {D.}~\bibnamefont {Pleiter}}, \bibinfo {author} {\bibfnamefont {P.~E.~L.}\ \bibnamefont {Rakow}}, \bibinfo {author} {\bibfnamefont {A.}~\bibnamefont {Schafer}}, \bibinfo {author} {\bibfnamefont {G.}~\bibnamefont {Schierholz}},\ and\ \bibinfo {author} {\bibfnamefont {W.}~\bibnamefont {Schroers}} (\bibinfo {collaboration} {QCDSF}),\ }\bibfield  {title} {\bibinfo {title} {{Generalized parton distributions from lattice QCD}},\ }\href {https://doi.org/10.1103/PhysRevLett.92.042002} {\bibfield  {journal} {\bibinfo  {journal} {Phys. Rev. Lett.}\ }\textbf {\bibinfo {volume} {92}},\ \bibinfo {pages} {042002} (\bibinfo {year} {2004})},\ \Eprint {https://arxiv.org/abs/hep-ph/0304249} {arXiv:hep-ph/0304249} \BibitemShut {NoStop}%
\bibitem [{\citenamefont {Alexandrou}\ \emph {et~al.}(2013)\citenamefont {Alexandrou}, \citenamefont {Constantinou}, \citenamefont {Dinter}, \citenamefont {Drach}, \citenamefont {Jansen}, \citenamefont {Kallidonis},\ and\ \citenamefont {Koutsou}}]{Alexandrou:2013joa}%
  \BibitemOpen
  \bibfield  {author} {\bibinfo {author} {\bibfnamefont {C.}~\bibnamefont {Alexandrou}}, \bibinfo {author} {\bibfnamefont {M.}~\bibnamefont {Constantinou}}, \bibinfo {author} {\bibfnamefont {S.}~\bibnamefont {Dinter}}, \bibinfo {author} {\bibfnamefont {V.}~\bibnamefont {Drach}}, \bibinfo {author} {\bibfnamefont {K.}~\bibnamefont {Jansen}}, \bibinfo {author} {\bibfnamefont {C.}~\bibnamefont {Kallidonis}},\ and\ \bibinfo {author} {\bibfnamefont {G.}~\bibnamefont {Koutsou}},\ }\bibfield  {title} {\bibinfo {title} {{Nucleon form factors and moments of generalized parton distributions using $N_f=2+1+1$ twisted mass fermions}},\ }\href {https://doi.org/10.1103/PhysRevD.88.014509} {\bibfield  {journal} {\bibinfo  {journal} {Phys. Rev. D}\ }\textbf {\bibinfo {volume} {88}},\ \bibinfo {pages} {014509} (\bibinfo {year} {2013})},\ \Eprint {https://arxiv.org/abs/1303.5979} {arXiv:1303.5979 [hep-lat]} \BibitemShut {NoStop}%
\bibitem [{\citenamefont {Deka}\ \emph {et~al.}(2015)\citenamefont {Deka} \emph {et~al.}}]{Deka:2013zha}%
  \BibitemOpen
  \bibfield  {author} {\bibinfo {author} {\bibfnamefont {M.}~\bibnamefont {Deka}} \emph {et~al.},\ }\bibfield  {title} {\bibinfo {title} {{Lattice study of quark and glue momenta and angular momenta in the nucleon}},\ }\href {https://doi.org/10.1103/PhysRevD.91.014505} {\bibfield  {journal} {\bibinfo  {journal} {Phys. Rev. D}\ }\textbf {\bibinfo {volume} {91}},\ \bibinfo {pages} {014505} (\bibinfo {year} {2015})},\ \Eprint {https://arxiv.org/abs/1312.4816} {arXiv:1312.4816 [hep-lat]} \BibitemShut {NoStop}%
\bibitem [{\citenamefont {Aoki}\ \emph {et~al.}(2020)\citenamefont {Aoki} \emph {et~al.}}]{FlavourLatticeAveragingGroup:2019iem}%
  \BibitemOpen
  \bibfield  {author} {\bibinfo {author} {\bibfnamefont {S.}~\bibnamefont {Aoki}} \emph {et~al.} (\bibinfo {collaboration} {Flavour Lattice Averaging Group}),\ }\bibfield  {title} {\bibinfo {title} {{FLAG Review 2019: Flavour Lattice Averaging Group (FLAG)}},\ }\href {https://doi.org/10.1140/epjc/s10052-019-7354-7} {\bibfield  {journal} {\bibinfo  {journal} {Eur. Phys. J. C}\ }\textbf {\bibinfo {volume} {80}},\ \bibinfo {pages} {113} (\bibinfo {year} {2020})},\ \Eprint {https://arxiv.org/abs/1902.08191} {arXiv:1902.08191 [hep-lat]} \BibitemShut {NoStop}%
\bibitem [{\citenamefont {Thomas}(2008)}]{Thomas:2008ga}%
  \BibitemOpen
  \bibfield  {author} {\bibinfo {author} {\bibfnamefont {A.~W.}\ \bibnamefont {Thomas}},\ }\bibfield  {title} {\bibinfo {title} {{Interplay of Spin and Orbital Angular Momentum in the Proton}},\ }\href {https://doi.org/10.1103/PhysRevLett.101.102003} {\bibfield  {journal} {\bibinfo  {journal} {Phys. Rev. Lett.}\ }\textbf {\bibinfo {volume} {101}},\ \bibinfo {pages} {102003} (\bibinfo {year} {2008})},\ \Eprint {https://arxiv.org/abs/0803.2775} {arXiv:0803.2775 [hep-ph]} \BibitemShut {NoStop}%
\bibitem [{\citenamefont {Altenbuchinger}\ \emph {et~al.}(2011)\citenamefont {Altenbuchinger}, \citenamefont {Hagler}, \citenamefont {Weise},\ and\ \citenamefont {Henley}}]{Altenbuchinger:2010sz}%
  \BibitemOpen
  \bibfield  {author} {\bibinfo {author} {\bibfnamefont {M.}~\bibnamefont {Altenbuchinger}}, \bibinfo {author} {\bibfnamefont {P.}~\bibnamefont {Hagler}}, \bibinfo {author} {\bibfnamefont {W.}~\bibnamefont {Weise}},\ and\ \bibinfo {author} {\bibfnamefont {E.~M.}\ \bibnamefont {Henley}},\ }\bibfield  {title} {\bibinfo {title} {{Spin structure of the nucleon: QCD evolution, lattice results and models}},\ }\href {https://doi.org/10.1140/epja/i2011-11140-2} {\bibfield  {journal} {\bibinfo  {journal} {Eur. Phys. J. A}\ }\textbf {\bibinfo {volume} {47}},\ \bibinfo {pages} {140} (\bibinfo {year} {2011})},\ \Eprint {https://arxiv.org/abs/1012.4409} {arXiv:1012.4409 [hep-ph]} \BibitemShut {NoStop}%
\bibitem [{\citenamefont {Ovchinnikov}\ \emph {et~al.}(1991)\citenamefont {Ovchinnikov}, \citenamefont {Pivovarov},\ and\ \citenamefont {Surguladze}}]{Ovchinnikov:1991mu}%
  \BibitemOpen
  \bibfield  {author} {\bibinfo {author} {\bibfnamefont {A.~A.}\ \bibnamefont {Ovchinnikov}}, \bibinfo {author} {\bibfnamefont {A.~A.}\ \bibnamefont {Pivovarov}},\ and\ \bibinfo {author} {\bibfnamefont {L.~R.}\ \bibnamefont {Surguladze}},\ }\bibfield  {title} {\bibinfo {title} {{Baryonic sum rules in the next-to-leading order in alpha-s}},\ }\href {https://doi.org/10.1142/S0217751X91001015} {\bibfield  {journal} {\bibinfo  {journal} {Int. J. Mod. Phys. A}\ }\textbf {\bibinfo {volume} {6}},\ \bibinfo {pages} {2025} (\bibinfo {year} {1991})}\BibitemShut {NoStop}%
\bibitem [{\citenamefont {Myhrer}\ and\ \citenamefont {Thomas}(2008)}]{Myhrer:2007cf}%
  \BibitemOpen
  \bibfield  {author} {\bibinfo {author} {\bibfnamefont {F.}~\bibnamefont {Myhrer}}\ and\ \bibinfo {author} {\bibfnamefont {A.~W.}\ \bibnamefont {Thomas}},\ }\bibfield  {title} {\bibinfo {title} {{A possible resolution of the proton spin problem}},\ }\href {https://doi.org/10.1016/j.physletb.2008.04.034} {\bibfield  {journal} {\bibinfo  {journal} {Phys. Lett. B}\ }\textbf {\bibinfo {volume} {663}},\ \bibinfo {pages} {302} (\bibinfo {year} {2008})},\ \Eprint {https://arxiv.org/abs/0709.4067} {arXiv:0709.4067 [hep-ph]} \BibitemShut {NoStop}%
\end{thebibliography}%

\end{document}